\documentclass[aip,rsi,amsmath,amssymb,preprint]{revtex4-1}

\usepackage{xspace}
\usepackage[usenames,dvipsnames]{color}
\usepackage{amsmath}
\usepackage{graphicx}
\usepackage{float}
\usepackage{lipsum,color,float}
\usepackage{subfig}
\usepackage{tikz}
\usepackage{siunitx}
\usepackage{tikz,pgfplots}
\usetikzlibrary{shapes,arrows,shapes.multipart,positioning,decorations.text,fit}
\usepackage{lipsum}
\usepackage{psfrag}
\usepackage{pgfplots}
\usepackage{color}

\pgfplotsset{compat=1.14}
\begin{document}
\title{ Robust procedure for creating and characterizing the atomic structure of scanning tunneling microscope tips}

\author{Sumit Tewari} 
\affiliation{Huygens-Kamerlingh Onnes Laboratory, Leiden University, Niels Bohrweg 2, 2333 CA Leiden, The Netherlands}
\author{Koen M. Bastiaans}
\affiliation{Huygens-Kamerlingh Onnes Laboratory, Leiden University, Niels Bohrweg 2, 2333 CA Leiden, The Netherlands} 
\author{Milan P. Allan}
\affiliation{Huygens-Kamerlingh Onnes Laboratory, Leiden University, Niels Bohrweg 2, 2333 CA Leiden, The Netherlands} 
\author{Jan M. van Ruitenbeek}
\affiliation{Huygens-Kamerlingh Onnes Laboratory, Leiden University, Niels Bohrweg 2, 2333 CA Leiden, The Netherlands}

\begin{abstract}
Scanning tunneling microscopes (STM) are used extensively for studying and manipulating matter at the atomic scale. In spite of the critical role of the STM tip, the control of the atomic-scale shape of STM tips remains a poorly solved problem. Here, we present a method for preparing tips {\it in-situ} and for ensuring the crystalline structure and reproducibly preparing tip structure up to the second atomic layer. We demonstrate a controlled evolution of such tips starting from undefined tip shapes.
\end{abstract}

\maketitle

\section{Introduction}
After the advent of scanning tunneling microscope (STM) in 1981 \cite{Binnig1982,Binnig1987}, it became possible to image conducting surfaces with atomic resolution. STM operates by bringing the apex of a fine metallic wire into tunneling distance from a surface of interest. By providing feedback in the tunnel current and scanning the tip over the surface one can make topographic maps of the surface with atomic resolution. STM has found its applications in many fields of science. Apart from studying surface topography, STM has been used for manipulating single atoms \cite{Eigler1990,Stroscio1991,Saw2014}, for doing spectroscopy \cite{wiesendanger1994}, for fabricating nano-structures with new tweaked electronic properties\cite{Manoharan2000}, for studying surface chemistry\cite{Frenken2017}, for probing collective \cite{Weitering1999} and local \cite{Pan2000} electronic behavior, and much more. 

It has been long known that the performance of the STM in these different fields of application is sensitive to the structure of the tip used in the measurements, since the signal is controlled by the overlaps of tip and surface electronic wave functions. Many techniques are available for preparing atomically clean sample surfaces, including repeated cycles of sputtering and annealing in case of metals, by fresh cleavage in case of suitable materials, and by high-temperature annealing, as for many semiconductors. However, a well defined tip structure at the atomic scale is still hard to achieve. Mechanical grinding \cite{Binnig1982}, electro-polishing \cite{Kojima1988} or electrochemical etching \cite{Heben1988, Victor2015} are standard {\it ex-situ} methods for preparing microscopically sharp tips. The tip apex can be cleaned {\it in-situ} using, e.g.,  Ar ion sputtering or electron bombardment \cite{Ernst2007}, but this may disrupt crystalline structure, which cannot be repaired by annealing since this renders a blunt tip. For all these methods, at the atomic scale the tip structure is poorly controlled and could even have multiple local apexes. For many purposes this does not hamper STM operation, since the tunnel current decays exponentially with tunnel gap so that the atom closest to the surface will dominate the imaging signal. However, the reproducible shape of current-voltage spectra depends strongly on the tip shape. It has been also known \cite{Ludoph2000} that the atomic structure behind the tip apex and the position of defects cause strong variations in current-voltage spectra. Controlled manipulation by STM of ad-atoms and molecules over metal surfaces depends also on the precise knowledge and reproducibility of the atomic tip structure. In the field of molecular electronics, where researchers are now trying to connect single molecules between an STM tip and a flat metal surface \cite{Wagner2015}, knowledge of tip shape is crucial.

Chen \cite{Chen1990} has shown that contrast of STM images depends on the choice of orbitals at the tip apex, explaining why decorating the tip apex with small molecules such as CO \cite{Chiang2014} has a large impact on image quality. These tip decorations have been well studied, but very little has been done in controlling the actual tip structure itself at the atomic scale. A first attempt in this direction was made in the pioneering article of Binning and Rohrer \cite{Binnig1982}, where they first introduced STM. They observed that the lateral resolution of their images can be increased when they gently touch the surface with the tip and then retract, which they describe as 'mini-spot-welding'.
In non-contact AFM experiments with specialized q-plus sensors, it has been shown \cite{Joachim2012,Joachim2013} that by making force gradient images of a CO molecule over Cu(111) substrate with a tungsten tip one obtains the information of the angular orbital symmetry of the front atom of the tip apex.
However, very few studies have been made to image the tip structure itself in-situ in STM and show its evolution upon tip preparation. 

Inspiration for our current approach comes from work with low-temperature STM under cryogenic vacuum \cite{Sabater2012}, where the authors report to have obtained crystalline tips by repeated deep indentation of a Au tip into a Au surface followed by retraction until the contact breaks. These indentation cycles cause plastic deformation of the tip apex \cite{Agrait1994}, which first gives random conductance traces but gradually evolves to repeatable cycles, which is interpreted as evidence for a crystalline tip structure.
This work on mechanical annealing was inspired by earlier break junction experiments, supported by molecular dynamics simulations  \cite{Agrait2003,Sabater2012}. 
A first application of this approach for a Au STM tip over a graphene surface \cite{Andres2012} was made by first locally depositing Au on the graphene surface from the tip using a high electric field pulse, followed by mechanical annealing similar to \cite{Sabater2012} over that Au deposit. The authors confirmed that the method improved the topographic contrast of the surface and the quality of the spectroscopic data.

Although this technique clearly holds promise for improvement of tip shaping and characterization, the potential has not been properly evaluated, which probably explains why it has not received more attention in the STM literature. Below we demonstrate a step by step evolution of the tip shape with a direct detection technique rather than studying conductance versus tip displacement or STM image contrast. 

\section{Experimental setup}
The experiments were performed in a Unisoku ultra high vacuum (UHV) and low temperature STM with base temperature of 2K. The in-plane (XY) scan range is set by the properties of the tube piezo and is limited to $1.5~\mu$m.  The basic operations of the STM are done using a RHK R9 controller. This controller was coupled to a custom-build MATLAB program for the tip-annealing procedures. The sample used for the tests is a 300~nm thick Au film deposited over mica. The Au surface was prepared {\it in-situ} by several cycles of Argon sputtering at 1~keV, at $5\cdot10^{-5}$ mbar for 15 minutes and annealing at 600 K for 1 minute to ensure a clean crystalline surface. The STM tips used in the experiments were commercial Unisoku Platinum Iridium (PtIr) tips having a tip radius of less than 20~nm, as obtained by grinding and mechanical polishing. PtIr tips are among the most commonly used tips in STM experiments as opposed to Au tip used in reference \cite{Sabater2012}. We exploit the fact that the tips become covered with the Au sample material at the apex. We anticipate that the nanoscale tip-surface interaction will be dominated by Au at both sides, while the PtIr base provides better mechanical stability. All the measurements reported here were performed at temperatures between 2 and 4.2~K.

\section{Conductance traces}

\begin{figure}
 \subfloat{\includegraphics[width = 3in]{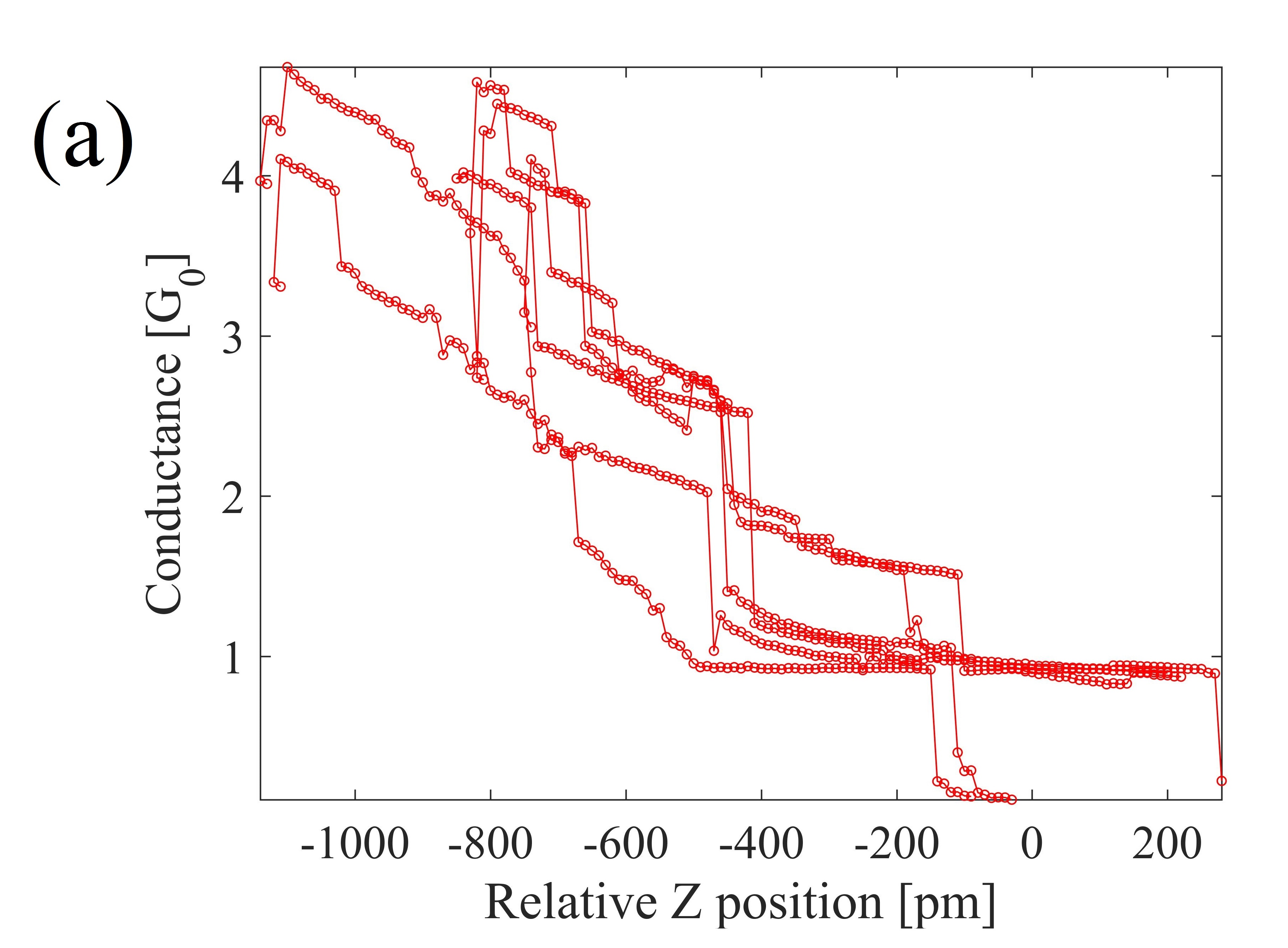}}\qquad 
 \subfloat{\includegraphics[width = 3in]{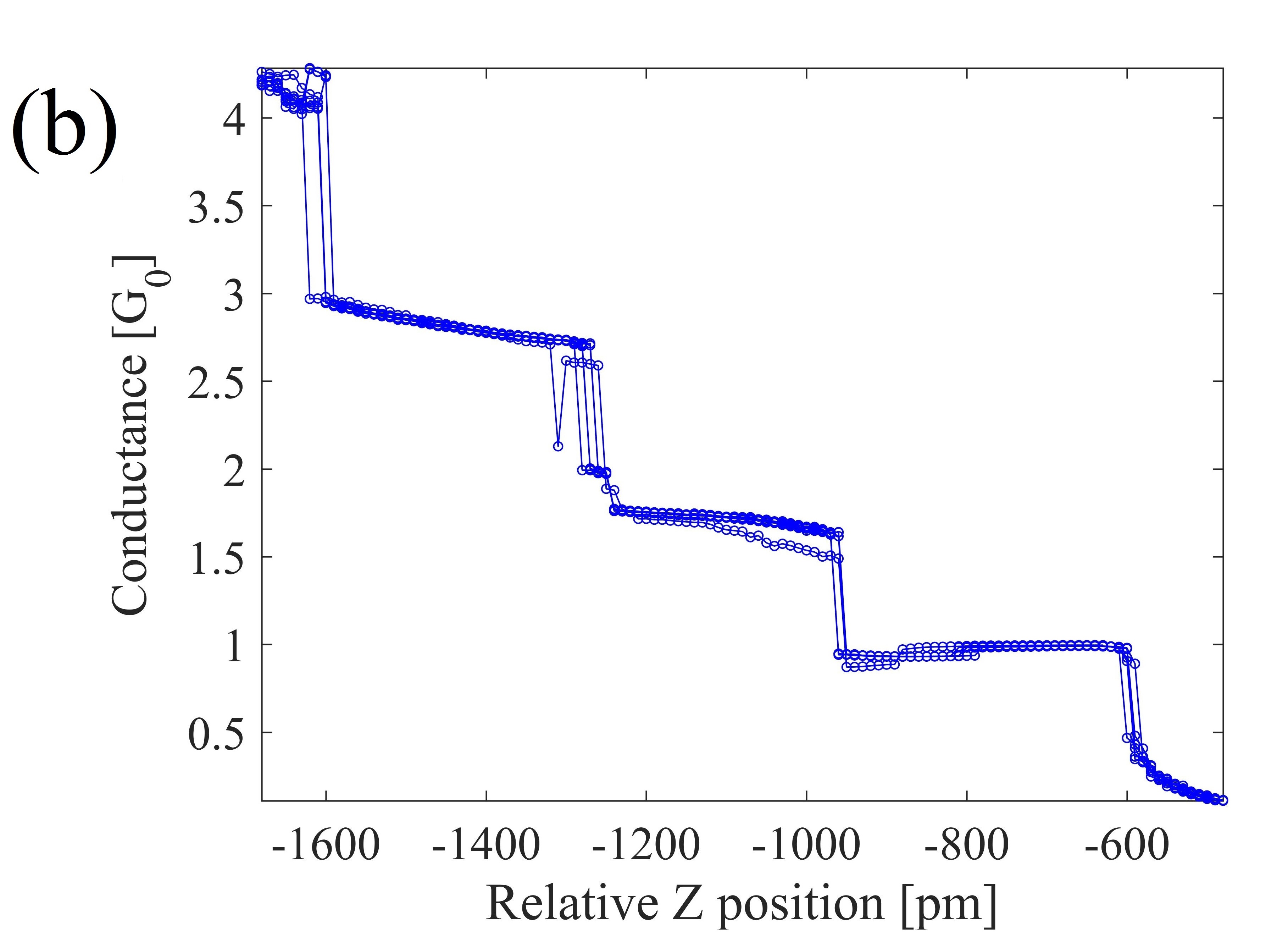}}
  \caption{Six consecutive conductance traces showing initially (a) non-repetitive structure, which is converted after mechanical annealing into (b) repetitive conductance traces. The conductance is expressed in units of  the quantum of conductance, $G_0 = 2e^2/h$. }
  \label{Fig1}
\end{figure}

Using this STM set-up we applied training procedures for the tip apex following the findings of Refs.~\cite{Sabater2012,Agrait1994}. To this purpose we set up the system to complete 450 mechanical contact annealing cycles and sampled the traces in groups of 10 cycles to probe for reproducibility in the traces. Figure~\ref{Fig1} shows examples of repetitive and non-repetitive cycles in our measurements. The system never settles permanently into either of the two cases. We see regular switching between repetitive and non-repetitive, which makes the maximum number of repetitive cycles of the order of 10-20. This appears to be different from the experiment by Sabater {\it et al.} \cite{Sabater2012} where, for a combination of Au tip and Au sample, the number of consecutively repeating traces was much larger. It has been reported that this repetition behavior during mechanical annealing is different for different materials \cite{Carlos2013}. The lower number of repeating conductance traces may be attributed to the fact that our tip is not purely Au, but PtIr covered with Au. In the experiments of Andres {\it et al.} \cite{Andres2012}, with a graphene surface covered with small cluster of Au the number of repeating traces was also limited, in their case to 16. 
 
\section{Procedure of tip preparation}
\begin{figure}
  \includegraphics[width=5.5in]{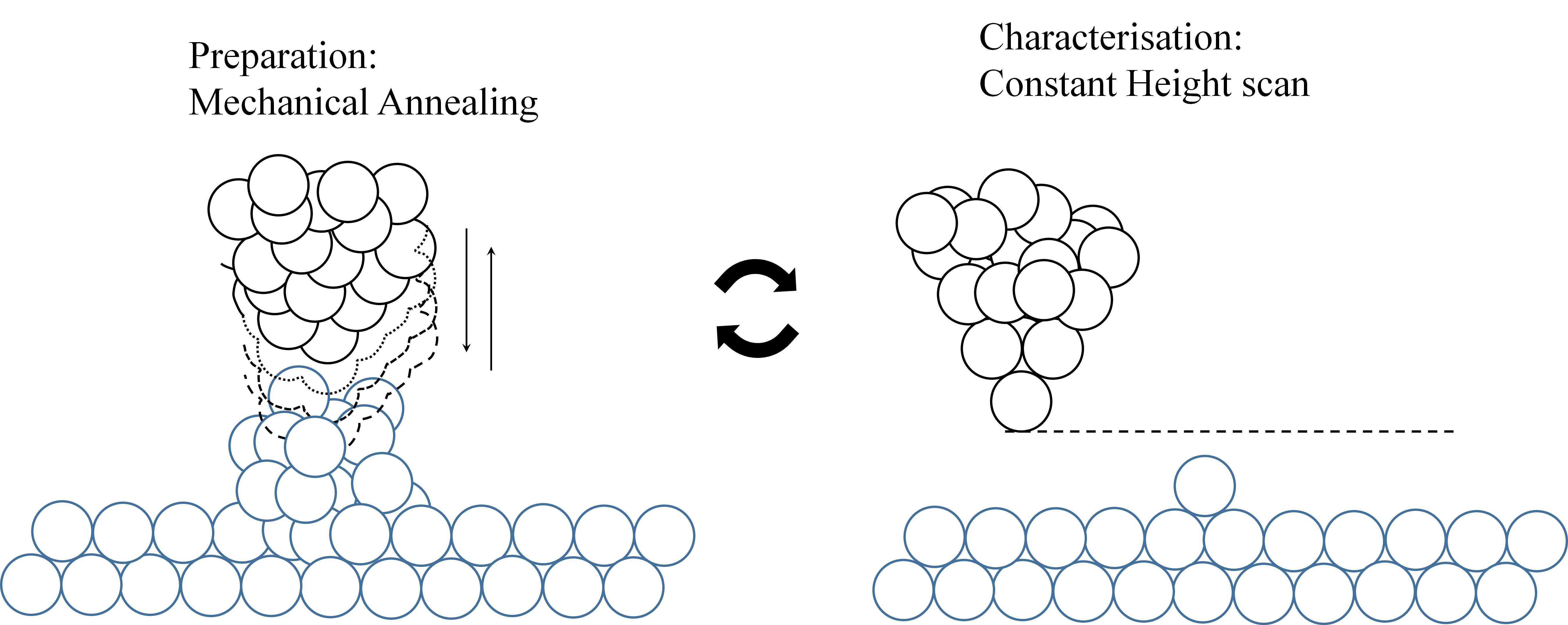}
  \caption{Schematic representation of the tip preparation process proposed in this article, where mechanical annealing cycles were followed by constant height scans made over a single ad-atom deposited on the surface.}
  \label{Fig0}
\end{figure}

We will now present a procedure, illustrated in Fig.~\ref{Fig0}, based upon this mechanical annealing that permits arriving at reproducible crystalline tip shapes starting from any random initial tip shape, and we show how we can verify the evolution of the tip shape. For this the STM tip is indented into the flat metal surface up to a pre-set conductance value and then retracted and this cycle is repeated many times. We then exploit the capabilities of  the low-temperature STM setup for imaging the evolution of the tip structure by scanning over an isolated ad-atom on the surface. The obtained STM images are convolutions of the topography and electronic states of sample and tip.  
Lang \cite{Lang1986} has shown theoretically using two planar metal electrodes with a single ad-atom on each of them, that upon scanning one with the other symmetric convolutions of the topography and electronic states are expected, giving circularly symmetric images. Any asymmetry in the STM images of an isolated ad-atom will reflect the asymmetry of the atomic structure behind the front atom.

\begin{figure}
\subfloat[]{\includegraphics[width = 1.8in]{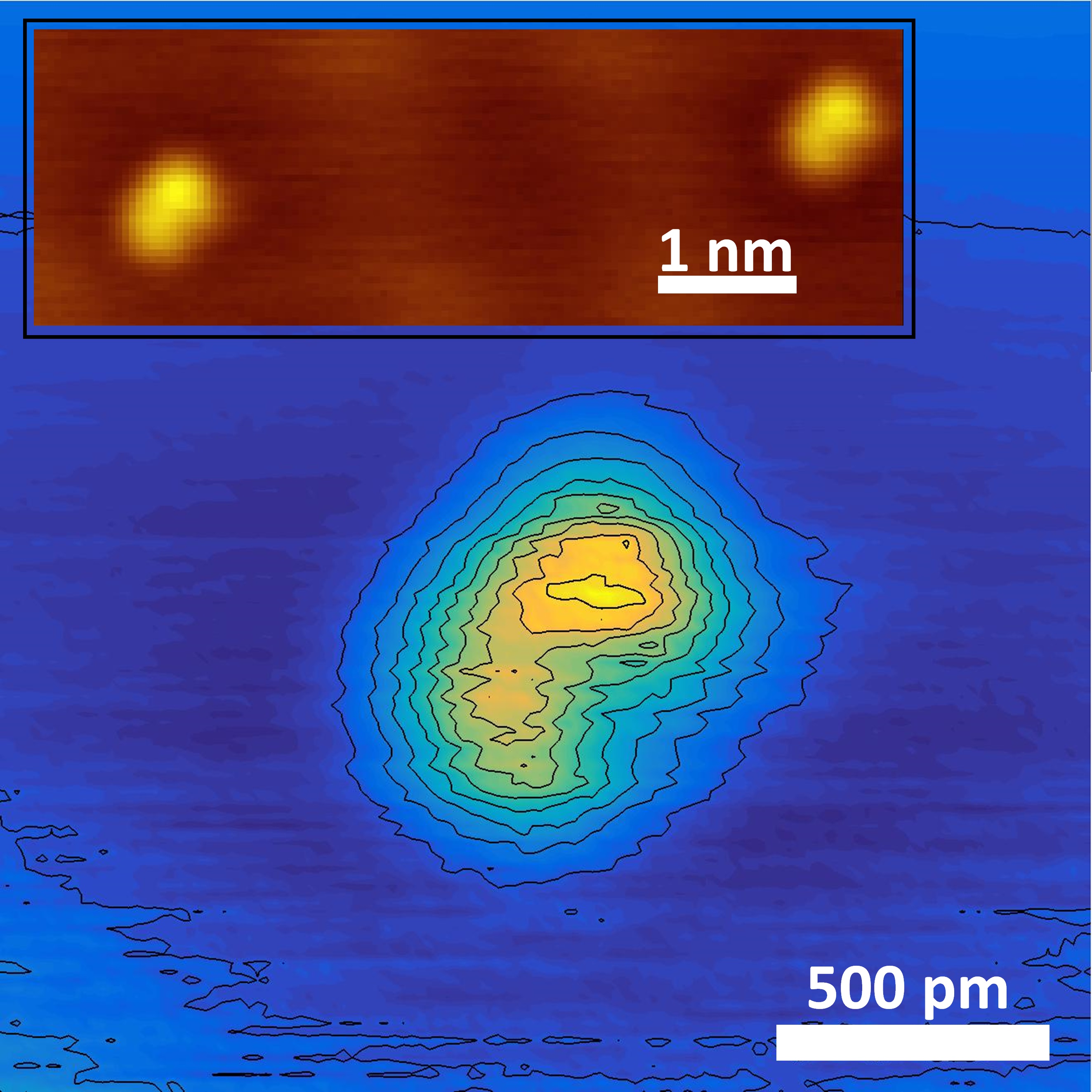}}\hspace*{\fill}
\subfloat[]{\includegraphics[width = 1.8in, height = 2.14in]{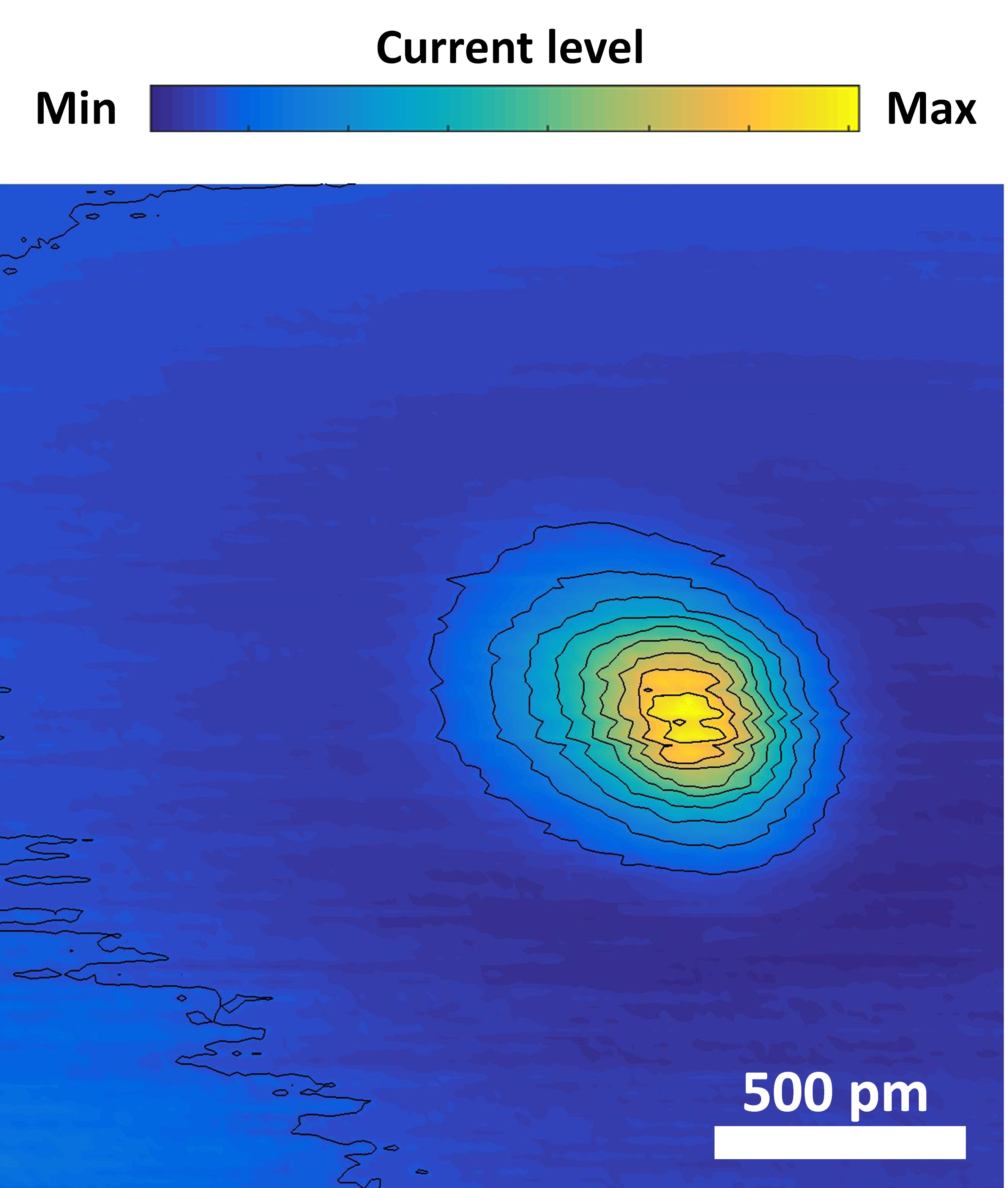}}\hspace*{\fill}
\subfloat[]{\includegraphics[width = 1.8in]{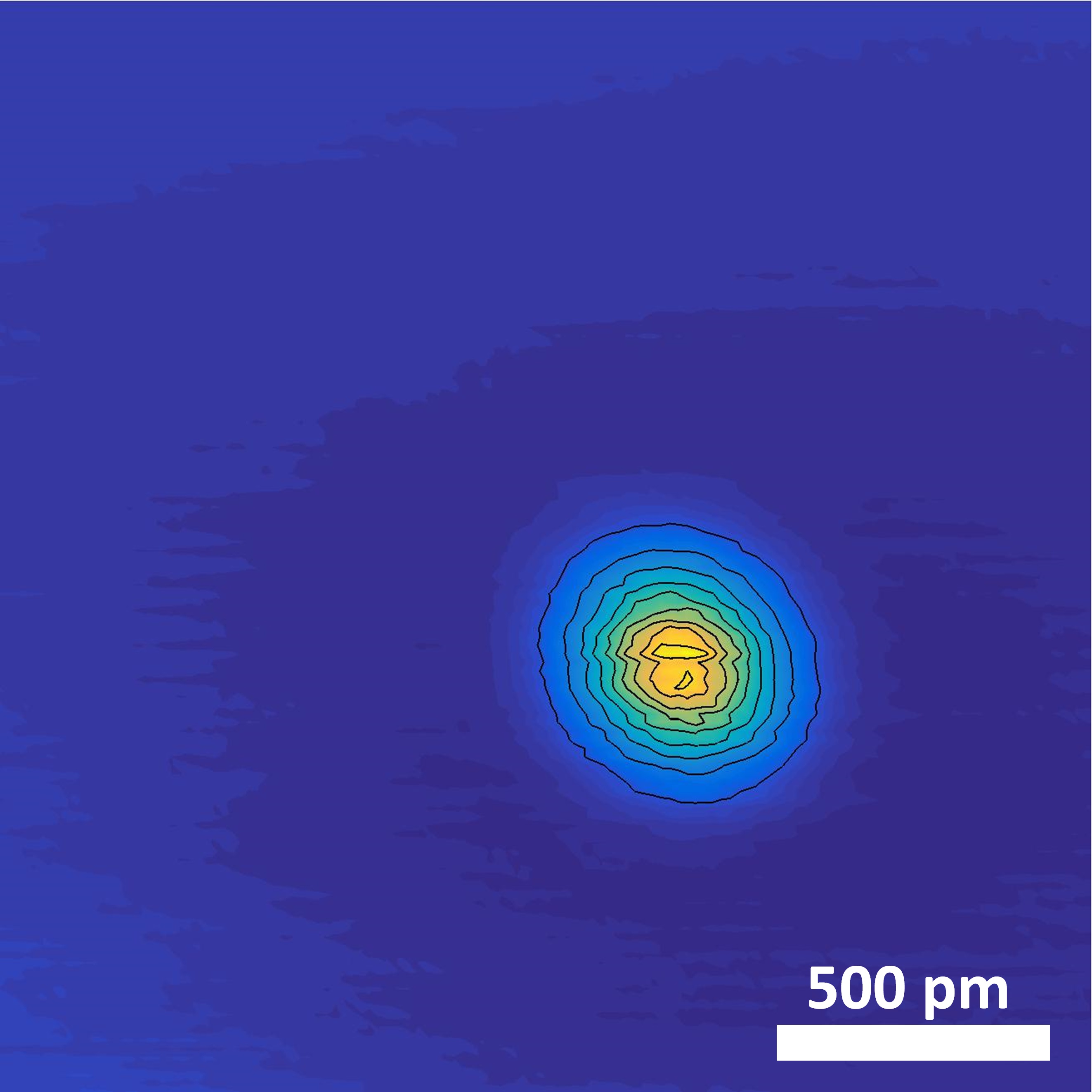}}\\
\subfloat[]{\includegraphics[width = 1.8in]{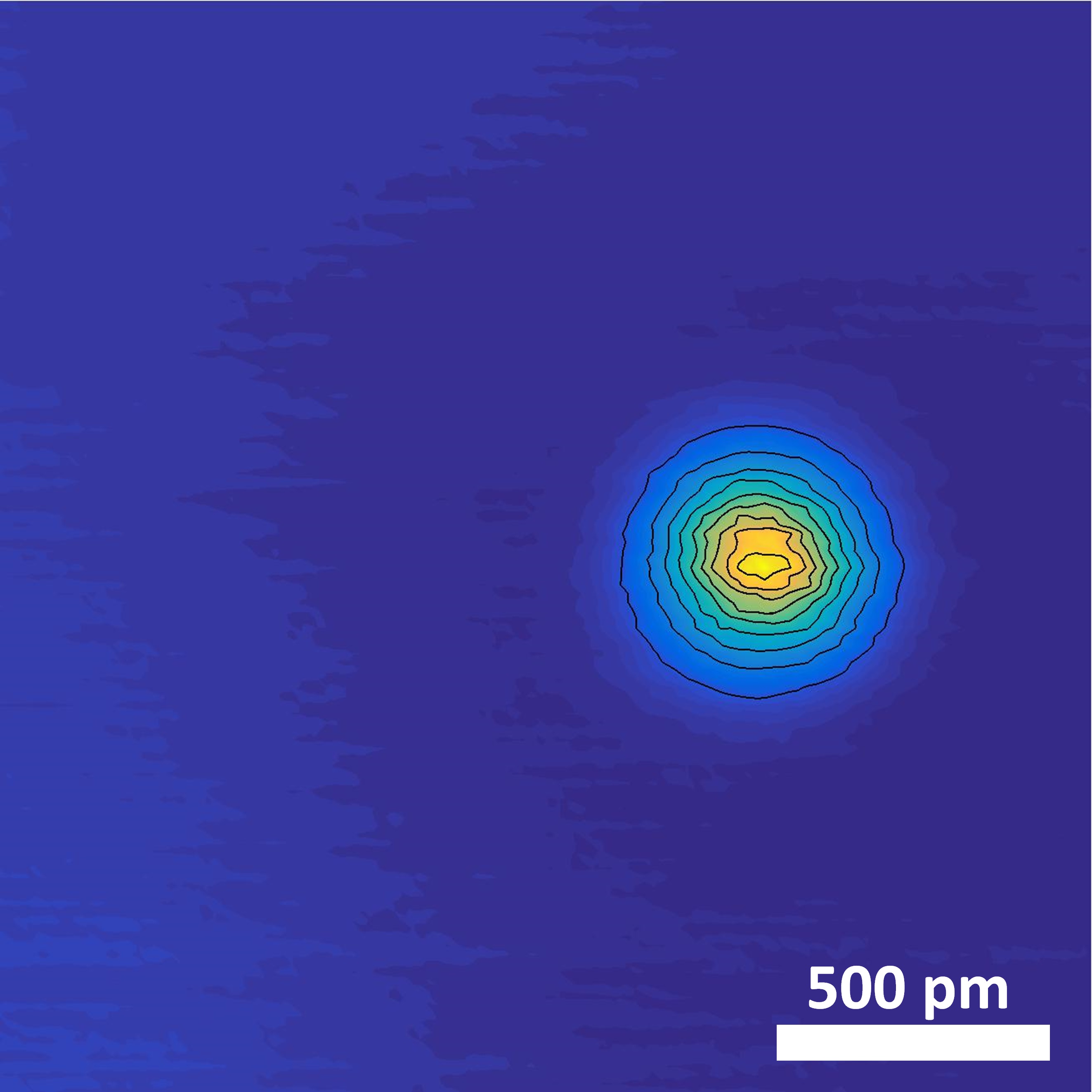}}\hspace*{\fill}
\subfloat[]{\includegraphics[width = 1.8in]{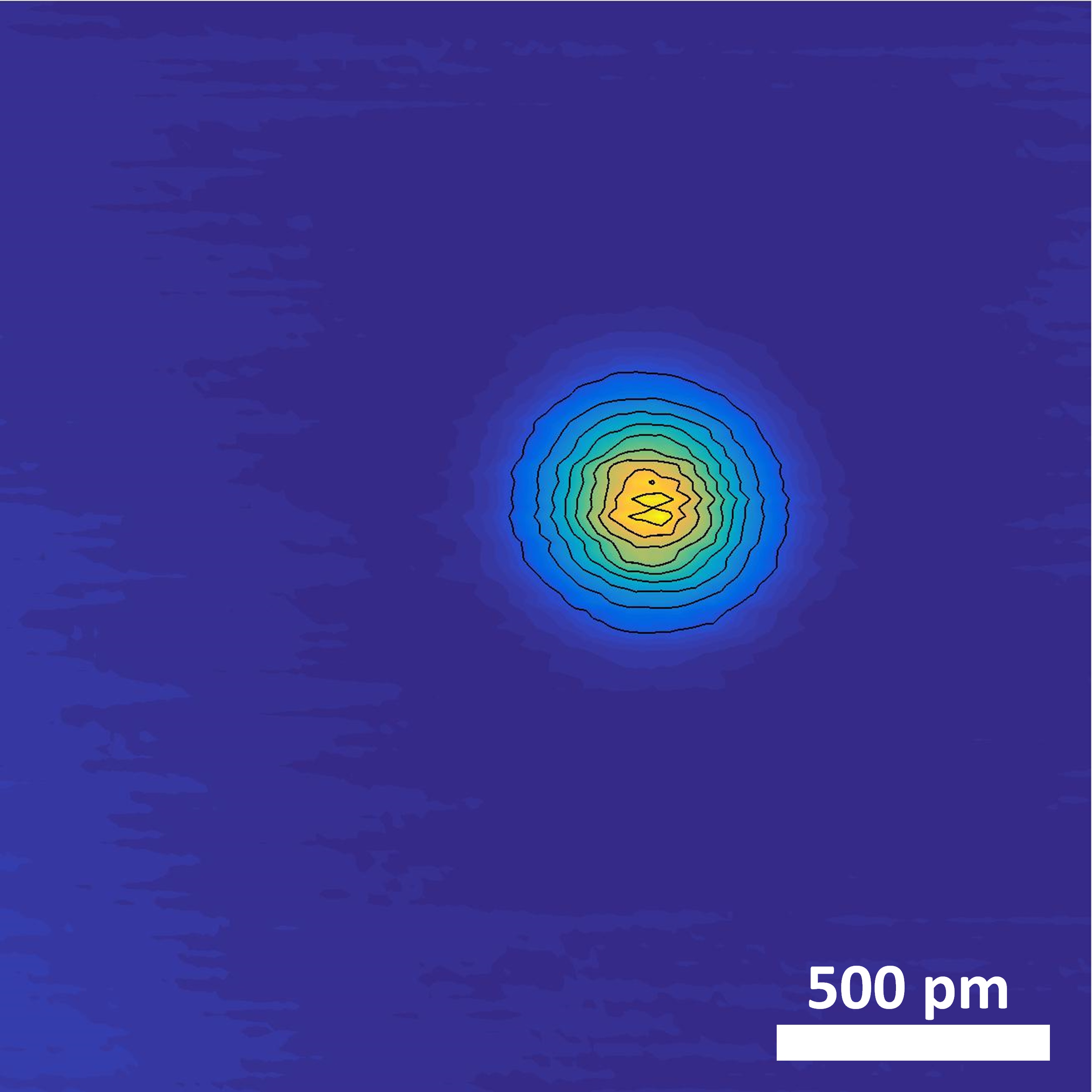}}\hspace*{\fill}
\subfloat[]{\includegraphics[width = 1.8in]{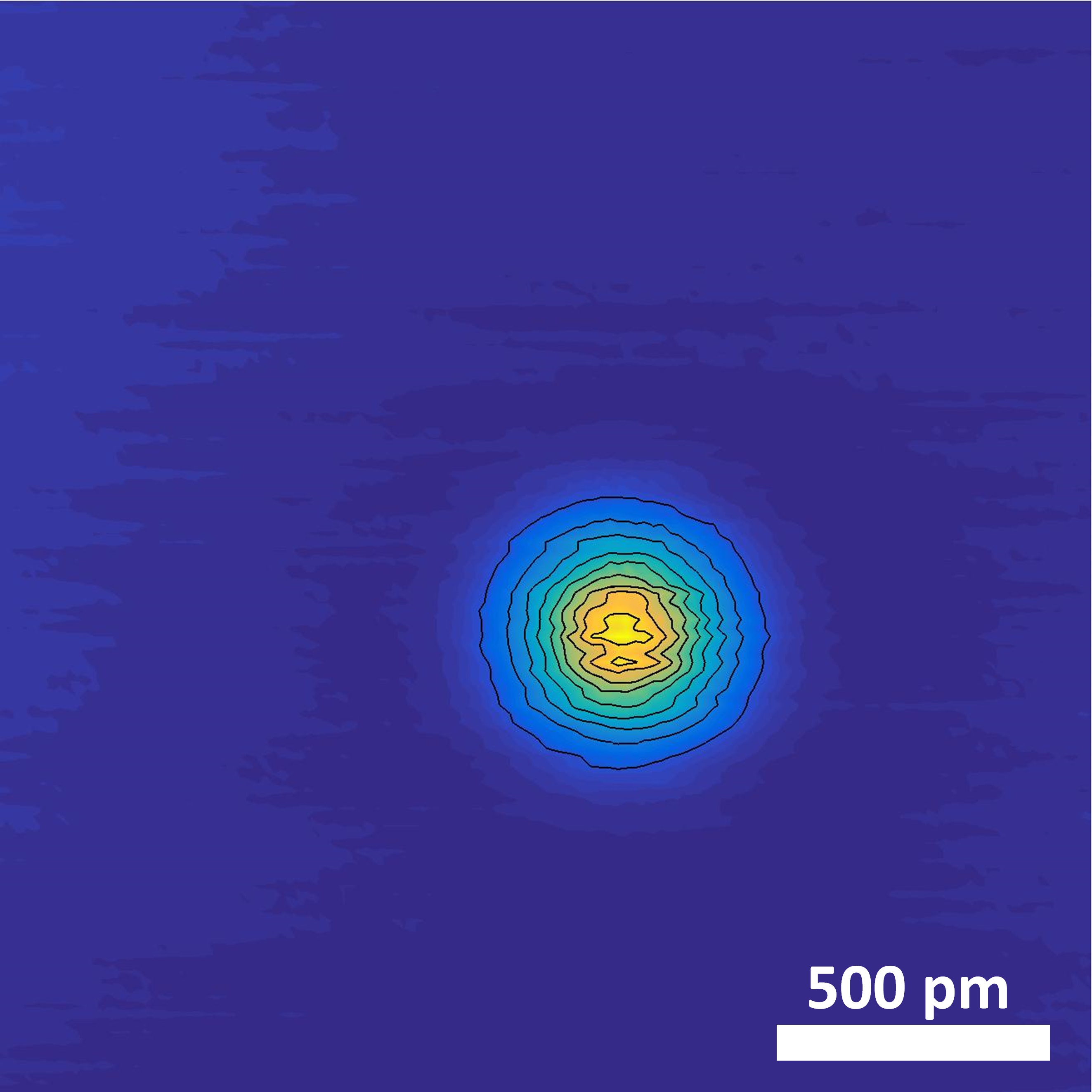}} 
\caption{The six panels show constant-height images of a single ad-atom. Here (a) shows a non-circular image of an ad-atom due to the random tip structure at the start. The inset of (a) shows constant-current image of two separate ad-atoms prepared on the surface to confirm that the asymmetric structure is due to the tip. The evolution of tip apex is shown in the next panels, leading towards a symmetric and reproducible structure (b-f) . Between each of the images we apply 20 mechanical annealing cycles. The contours shown are linearly spaced in current. For ease of comparison the current levels in the images have been normalized to the maximum current level, which for the panels (a)-(f) are 66~nA, 31~nA, 26~nA, 26~nA , 25~nA and 22~nA, respectively.}
\label{Fig3}
\end{figure}

We start the experiments by depositing a single ad-atom from the Au covered PtIr tip in the center of the FCC sector of the Au(111) herringbone reconstruction, following the procedure of Ref.~\cite{Saw2014}. 
 At 100~mV bias, once thermal and mechanical drift has been stabilized, we release the current feedback and move the tip towards surface at a rate of $0.5$~\AA/s  using a custom build program in MATLAB. The motion is stopped once the conductance reaches the quantum of conductance ($G_0 = 2e^2/h$, which is what we expect for a single-atom point contact in Au \cite{Agrait2003}),  followed by tip retraction back into the tunneling regime. This procedure leaves, with high success rate, a single Au ad-atom on the surface. The result is imaged in the usual topographic mode of STM shown as an inset in Fig.~\ref{Fig3} (a), which demonstrates that the ad-atom does not have circular symmetry. In order to verify that this asymmetry is associated with the tip, we deposit a second ad-atom. The image of the second atom is a replica of the first, in shape and orientation, confirming that the asymmetry is associated with the tip shape.  

In the next steps we use an individual ad-atom for imaging the structure of the tip apex, employing constant-height mode scans with a box size of 2~nm centered on the ad-atom. 
The goal is to pick up tunneling current signal from the second row of atoms above the apex atom. For FCC packing the distance to the second row is 2.5~\AA~ larger than that to the apex atom,  from which we estimate the current level to the second-row atoms at 100~pA for a current of 30~nA at the apex atom, which is well above the noise floor of 10~pA at a gain of $10^8$~V/A for the current amplifier. As the tunnel current varies exponentially with distance, even small deviations from surface-parallel FCC packing of the second-layer tip atoms will give a detectable contribution.
 In order to scan at such high tunnel current we first switch off the current feedback and bring the tip closer to the flat part of the surface to a fixed tunnel current value. This value is chosen to ensure that the current is as high as 30nA, when the tip is over the ad-atom during scan. Then we take the constant-height scan at a tip speed of 2~nm/s. Figure \ref{Fig3}(a) shows an example of the resulting image at the initial stage, before starting the tip preparation procedure. 

After imaging the tip apex we move the tip to an edge of the scan range, about 700~nm away from the ad-atom, and perform a series of 20 mechanical annealing cycles. For this we use the same MATLAB controlled procedure as described above, except that we indent the tip farther until it reaches a preset conductance of 4~$G_0$. After 20 of these mechanical annealing cycles, we return to the original ad-atom and image it again using same procedure as above. We repeat these steps of mechanical annealing and imaging until we obtain a symmetric tip image, as illustrated in Figs.~\ref{Fig3}(a-f).

\begin{figure}
\subfloat[start]{\includegraphics[width = 2in]{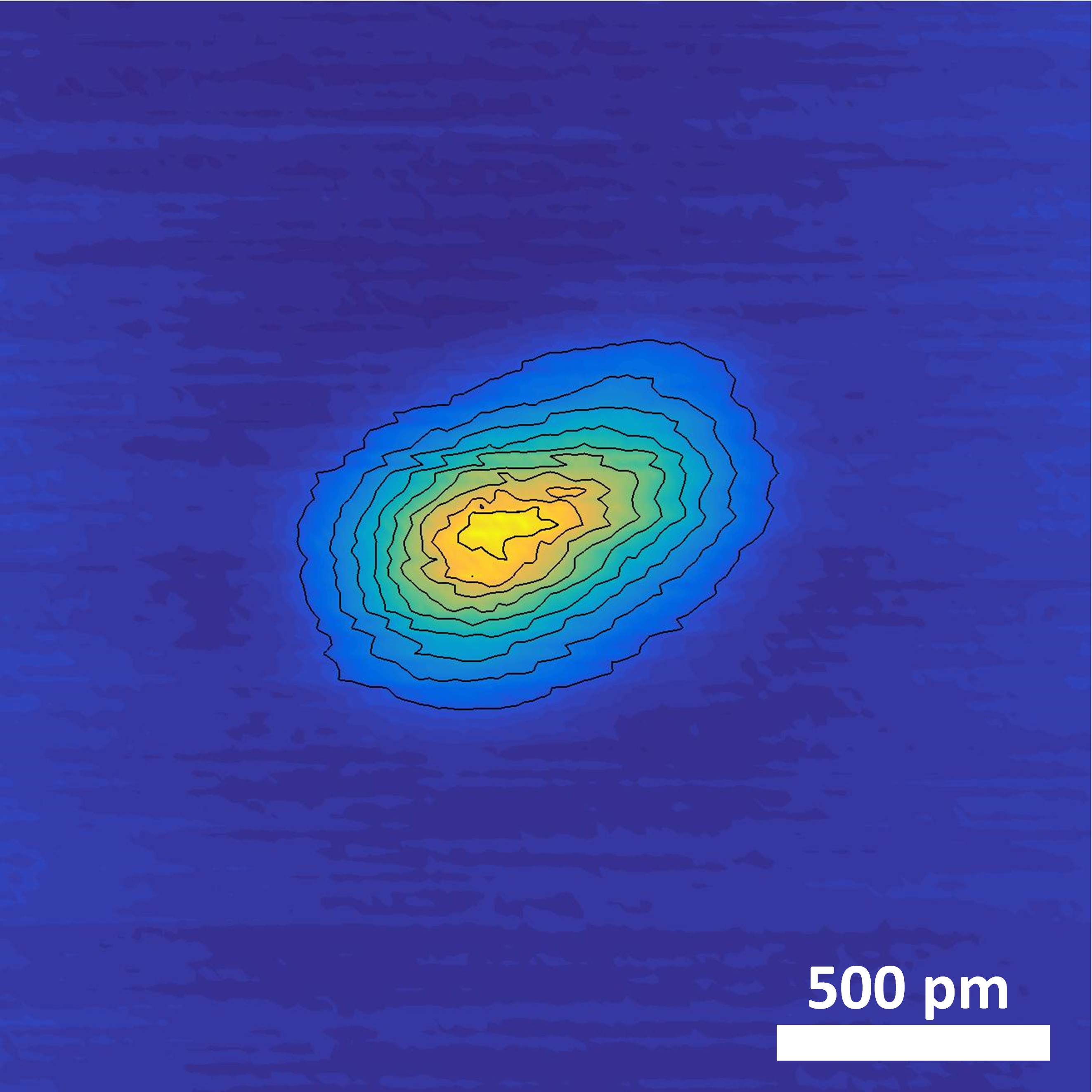}}\qquad 
\subfloat[end]{\includegraphics[width = 2.375in]{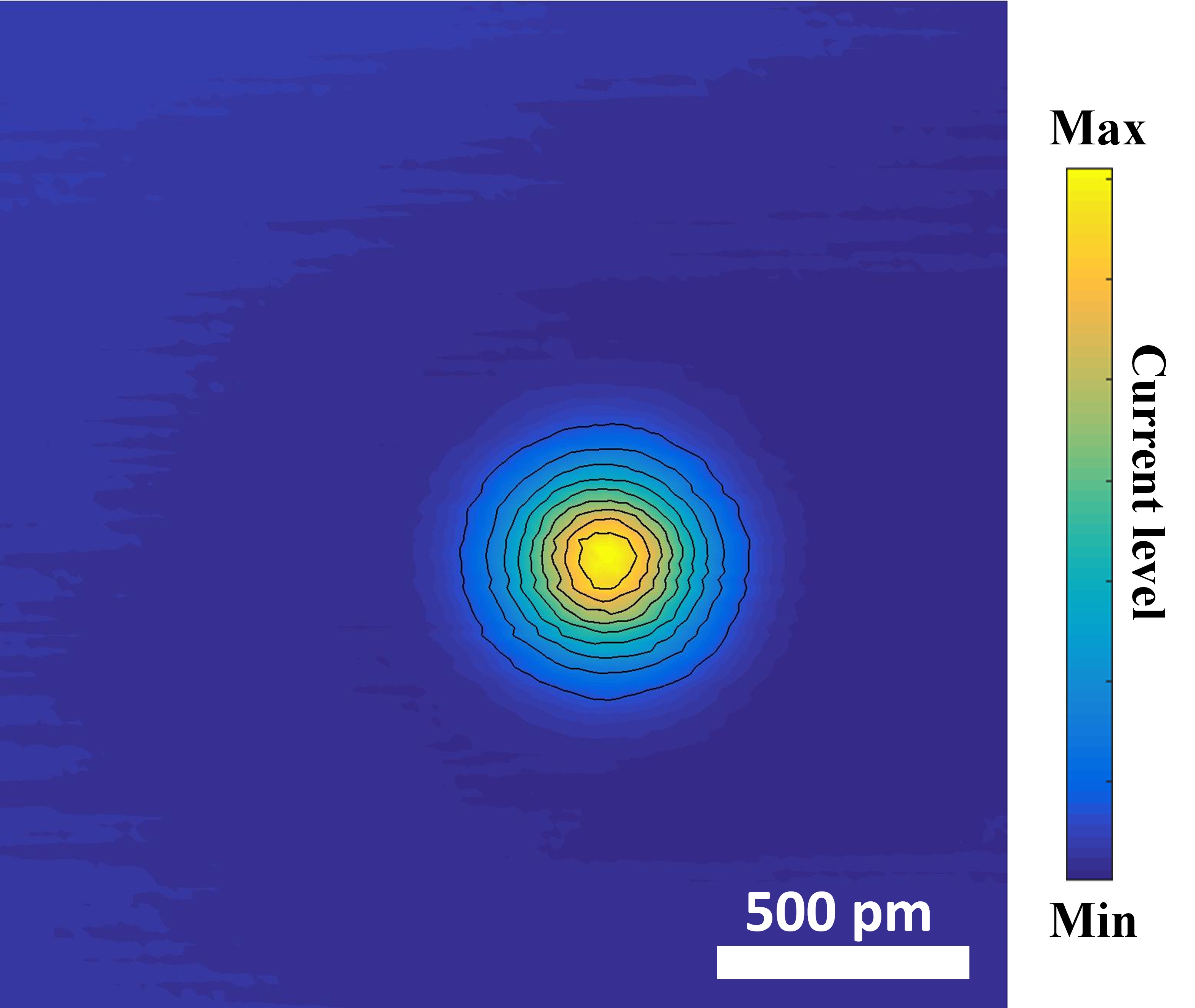}}
\caption{Normalized constant-height images starting from another initial tip apex (a) evolving towards a symmetric apex at the end of the procedure (b). The contours shown are linearly spaced in current. For ease of comparison the current levels in the images have been normalized to the maximum current level which for the above figures are 38~nA and 24~nA, respectively}
\label{Fig4}
\end{figure}

We have verified the reproducibility by repeating the procedure for other tips and other ad-atoms. Figure~\ref{Fig4} shows the initial and final stage for another run. Although we occasionally observe a jump back to a more asymmetric tip configuration, the typical behavior follows a smooth evolution towards a reproducible symmetric image.

\begin{figure}
\includegraphics[width = 4in]{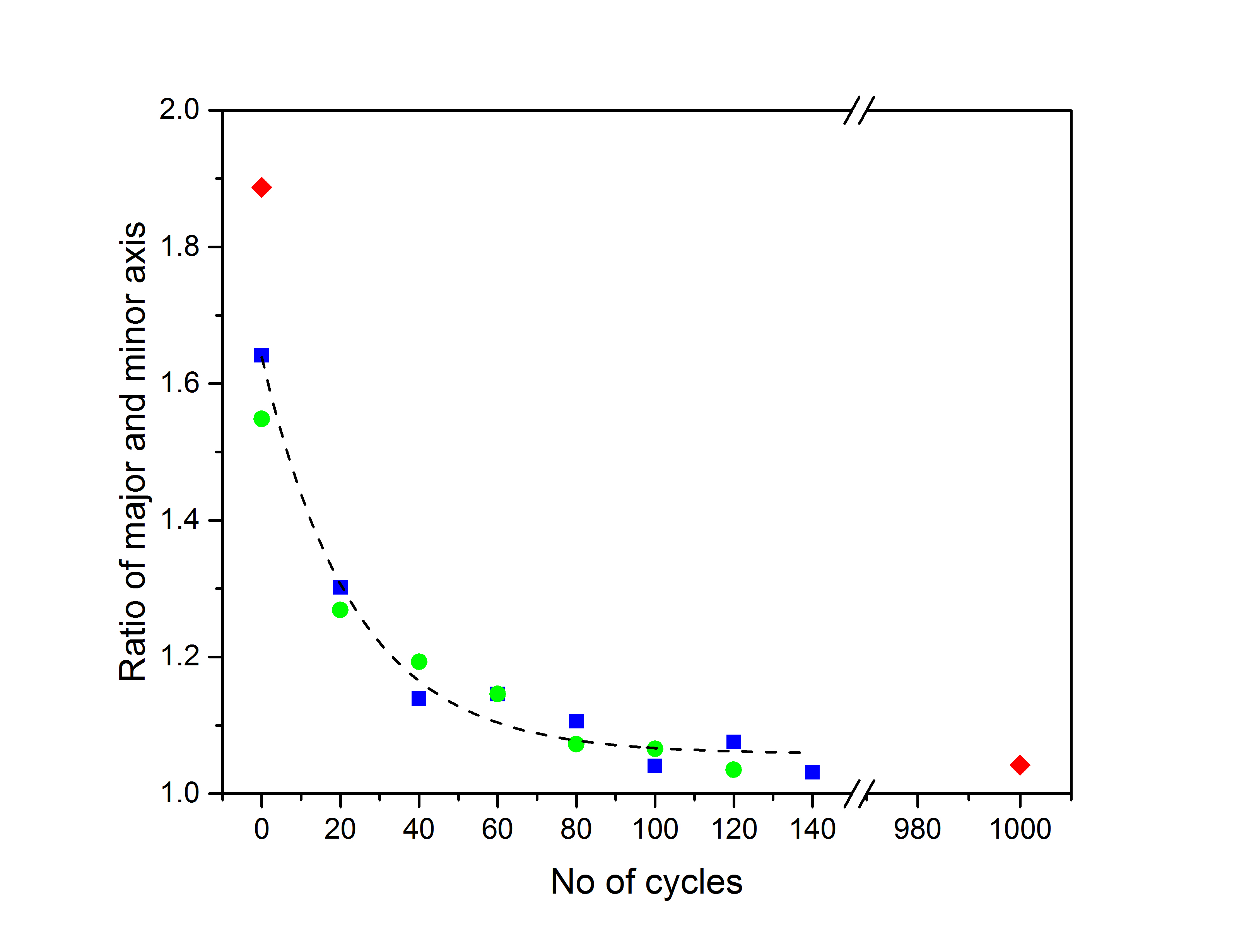}
\caption{Ratio of the major to minor axis of the ellipses fit showing convergence of the tip apex structure to a circularly symmetric shape. Three independent runs are shown by blue, green and red symbols. The two points shown as red diamonds represent initial and final images of an ad-atom for 1000 mechanical annealing cycles. The black dashed curve is a guide to the eye showing exponentially smooth transition to a circularly symmetric state.}
\label{Fig5}
\end{figure}

For analyzing the results of this procedure more quantitatively we would like to define a parameter which could capture the convergence of the tip structure to this symmetric state. This same parameter should also serve to verify whether both tips in  Figs.~\ref{Fig3} and \ref{Fig4} converged to the same state.  For comparing the constant height images we select a contour at a fixed tunnel current level (which for our case is 13.3 nA) above the background and fit an ellipse to this contour, which is meaningful as the tip was brought close to the surface to approximately the same tunnel current value before the start of each scan. From the fit we extract the ratio of major to minor axis of the ellipse, $a/b$, as a measure of the deviation from circular symmetry. Figure~\ref{Fig5} shows a plot of  the evolution of the $a/b$ ratio with the number of annealing cycles. The blue and the green data points show two independent runs, for different tips and different samples. The red dots show the start and end points for a third set of data, also shown in Fig.~\ref{Fig4}, applying 1000 mechanical annealing cycles. Surprisingly, the evolution of the tip is very regular and reproducible.  Starting from uncontrolled and asymmetric tip apex configurations, we find that the $a/b$ ratio decays following a common pattern, and arrives at similar minimum values. The deviation of about 4\%~ from 1 for $a/b$ ratio is probably limited by thermal drift and electrical noise of our system. The curve shows that the data are closely described by an exponential dependence. 

\section{Conclusion}
We have demonstrated a method for shaping a metallic tip apex in STM. By placing an ad-atom over a smooth Au surface the structure of the tip apex can be imaged, and we find that the shape of the STM tip evolves surprisingly smoothly and reproducibly towards an atomically sharp and symmetric structure of the second layer from the tip apex atom, starting from any random and poorly defined tip shapes. As has been illustrated by molecular dynamics simulations \cite{Sabater2012} the mechanical annealing cycles lead to a more regular atomic packing at both sides of the junction. The smooth evolution of this alignment observed here suggests that the process evolves through gradual shifts in packing and orientation of the layers farther away from the apex atom. 
Based on previous break junction experiments \cite{Agrait2003} it is known that for less than 100~mV bias voltage the joule heating of the junctions is negligible and so the current level during mechanical annealing should should not play an important role, when keeping the bias below this value. The depth of indentation is more critical: just touching the surface with the front atom is not enough, and much deeper indentation does not result in reproducible conductance cycles \cite{Carlos2013}. We have demonstrated the procedure here only for the combination of PtIr covered with Au tip and Au surface, but we expect that our method is not limited to this combination of tip and sample material. The procedure is simple to implement in any low-temperature STM operating under UHV, and will be very useful for increasing the reproducibility of imaging and spectroscopy. 

\section{Acknowledgment}
This work was supported by the Netherlands Organization for Scientic Research (NWO/OCW), as part of the Frontiers of Nanoscience program and the Vidi talent scheme (M.P.A.).


\begin{thebibliography}{26}%
\makeatletter
\providecommand \@ifxundefined [1]{%
 \@ifx{#1\undefined}
}%
\providecommand \@ifnum [1]{%
 \ifnum #1\expandafter \@firstoftwo
 \else \expandafter \@secondoftwo
 \fi
}%
\providecommand \@ifx [1]{%
 \ifx #1\expandafter \@firstoftwo
 \else \expandafter \@secondoftwo
 \fi
}%
\providecommand \natexlab [1]{#1}%
\providecommand \enquote  [1]{``#1''}%
\providecommand \bibnamefont  [1]{#1}%
\providecommand \bibfnamefont [1]{#1}%
\providecommand \citenamefont [1]{#1}%
\providecommand \href@noop [0]{\@secondoftwo}%
\providecommand \href [0]{\begingroup \@sanitize@url \@href}%
\providecommand \@href[1]{\@@startlink{#1}\@@href}%
\providecommand \@@href[1]{\endgroup#1\@@endlink}%
\providecommand \@sanitize@url [0]{\catcode `\\12\catcode `\$12\catcode
  `\&12\catcode `\#12\catcode `\^12\catcode `\_12\catcode `\%12\relax}%
\providecommand \@@startlink[1]{}%
\providecommand \@@endlink[0]{}%
\providecommand \url  [0]{\begingroup\@sanitize@url \@url }%
\providecommand \@url [1]{\endgroup\@href {#1}{\urlprefix }}%
\providecommand \urlprefix  [0]{URL }%
\providecommand \Eprint [0]{\href }%
\providecommand \doibase [0]{http://dx.doi.org/}%
\providecommand \selectlanguage [0]{\@gobble}%
\providecommand \bibinfo  [0]{\@secondoftwo}%
\providecommand \bibfield  [0]{\@secondoftwo}%
\providecommand \translation [1]{[#1]}%
\providecommand \BibitemOpen [0]{}%
\providecommand \bibitemStop [0]{}%
\providecommand \bibitemNoStop [0]{.\EOS\space}%
\providecommand \EOS [0]{\spacefactor3000\relax}%
\providecommand \BibitemShut  [1]{\csname bibitem#1\endcsname}%
\let\auto@bib@innerbib\@empty
\bibitem [{\citenamefont {Binnig}\ \emph {et~al.}(1982)\citenamefont {Binnig},
  \citenamefont {Rohrer}, \citenamefont {Gerber},\ and\ \citenamefont
  {Weibel}}]{Binnig1982}%
  \BibitemOpen
  \bibfield  {author} {\bibinfo {author} {\bibfnamefont {G.}~\bibnamefont
  {Binnig}}, \bibinfo {author} {\bibfnamefont {H.}~\bibnamefont {Rohrer}},
  \bibinfo {author} {\bibfnamefont {C.}~\bibnamefont {Gerber}}, \ and\ \bibinfo
  {author} {\bibfnamefont {E.}~\bibnamefont {Weibel}},\ }\href {\doibase
  10.1103/PhysRevLett.49.57} {\bibfield  {journal} {\bibinfo  {journal} {Phys.
  Rev. Lett.}\ }\textbf {\bibinfo {volume} {49}},\ \bibinfo {pages} {57}
  (\bibinfo {year} {1982})}\BibitemShut {NoStop}%
\bibitem [{\citenamefont {Binnig}\ and\ \citenamefont
  {Rohrer}(1987)}]{Binnig1987}%
  \BibitemOpen
  \bibfield  {author} {\bibinfo {author} {\bibfnamefont {G.}~\bibnamefont
  {Binnig}}\ and\ \bibinfo {author} {\bibfnamefont {H.}~\bibnamefont
  {Rohrer}},\ }\href {\doibase 10.1103/RevModPhys.59.615} {\bibfield  {journal}
  {\bibinfo  {journal} {Rev. Mod. Phys.}\ }\textbf {\bibinfo {volume} {59}},\
  \bibinfo {pages} {615} (\bibinfo {year} {1987})}\BibitemShut {NoStop}%
\bibitem [{\citenamefont {Eigler}\ and\ \citenamefont
  {Schweizer}(1990)}]{Eigler1990}%
  \BibitemOpen
  \bibfield  {author} {\bibinfo {author} {\bibfnamefont {D.~M.}\ \bibnamefont
  {Eigler}}\ and\ \bibinfo {author} {\bibfnamefont {E.~K.}\ \bibnamefont
  {Schweizer}},\ }\href {\doibase 10.1038/344524a0M3} {\bibfield  {journal}
  {\bibinfo  {journal} {Nature}\ }\textbf {\bibinfo {volume} {344}},\ \bibinfo
  {pages} {524} (\bibinfo {year} {1990})}\BibitemShut {NoStop}%
\bibitem [{\citenamefont {Stroscio}\ and\ \citenamefont
  {Eigler}(1991)}]{Stroscio1991}%
  \BibitemOpen
  \bibfield  {author} {\bibinfo {author} {\bibfnamefont {J.~A.}\ \bibnamefont
  {Stroscio}}\ and\ \bibinfo {author} {\bibfnamefont {D.~M.}\ \bibnamefont
  {Eigler}},\ }\href {\doibase 10.1126/science.254.5036.1319} {\bibfield
  {journal} {\bibinfo  {journal} {Science}\ }\textbf {\bibinfo {volume}
  {254}},\ \bibinfo {pages} {1319} (\bibinfo {year} {1991})},\ \Eprint
  {http://arxiv.org/abs/http://science.sciencemag.org/content/254/5036/1319.full.pdf}
  {http://science.sciencemag.org/content/254/5036/1319.full.pdf} \BibitemShut
  {NoStop}%
\bibitem [{\citenamefont {Hla}(2014)}]{Saw2014}%
  \BibitemOpen
  \bibfield  {author} {\bibinfo {author} {\bibfnamefont {S.~W.}\ \bibnamefont
  {Hla}},\ }\href {\doibase 10.1088/0034-4885/77/5/056502} {\bibfield
  {journal} {\bibinfo  {journal} {Reports on Progress in Physics}\ }\textbf
  {\bibinfo {volume} {77}},\ \bibinfo {pages} {056502} (\bibinfo {year}
  {2014})}\BibitemShut {NoStop}%
\bibitem [{\citenamefont {Wiesendanger}(1994)}]{wiesendanger1994}%
  \BibitemOpen
  \bibfield  {author} {\bibinfo {author} {\bibfnamefont {R.}~\bibnamefont
  {Wiesendanger}},\ }\href {\doibase 10.1017/CBO9780511524356} {\emph {\bibinfo
  {title} {Scanning Probe Microscopy and Spectroscopy: Methods and
  Applications}}},\ Scanning Probe Microscopy and Spectroscopy: Methods and
  Applications\ (\bibinfo  {publisher} {Cambridge University Press},\ \bibinfo
  {year} {1994})\BibitemShut {NoStop}%
\bibitem [{\citenamefont {Manoharan}, \citenamefont {Lutz},\ and\ \citenamefont
  {Eigler}(2000)}]{Manoharan2000}%
  \BibitemOpen
  \bibfield  {author} {\bibinfo {author} {\bibfnamefont {H.~C.}\ \bibnamefont
  {Manoharan}}, \bibinfo {author} {\bibfnamefont {C.~P.}\ \bibnamefont {Lutz}},
  \ and\ \bibinfo {author} {\bibfnamefont {D.~M.}\ \bibnamefont {Eigler}},\
  }\href {\doibase 10.1038/35000508M3} {\bibfield  {journal} {\bibinfo
  {journal} {Nature}\ }\textbf {\bibinfo {volume} {403}},\ \bibinfo {pages} {3}
  (\bibinfo {year} {2000})}\BibitemShut {NoStop}%
\bibitem [{\citenamefont {Frenken}\ and\ \citenamefont
  {Groot}(2017)}]{Frenken2017}%
  \BibitemOpen
  \bibfield  {author} {\bibinfo {author} {\bibfnamefont {J.}~\bibnamefont
  {Frenken}}\ and\ \bibinfo {author} {\bibfnamefont {I.}~\bibnamefont
  {Groot}},\ }\href {\doibase 10.1007/978-3-319-44439-0} {\emph {\bibinfo
  {title} {Operando Research in Heterogeneous Catalysis}}},\ Springer Series in
  Chemical Physics\ (\bibinfo  {publisher} {Springer International
  Publishing},\ \bibinfo {year} {2017})\BibitemShut {NoStop}%
\bibitem [{\citenamefont {Weitering}\ \emph {et~al.}(1999)\citenamefont
  {Weitering}, \citenamefont {Carpinelli}, \citenamefont {Melechko},
  \citenamefont {Zhang}, \citenamefont {Bartkowiak},\ and\ \citenamefont
  {Plummer}}]{Weitering1999}%
  \BibitemOpen
  \bibfield  {author} {\bibinfo {author} {\bibfnamefont {H.~H.}\ \bibnamefont
  {Weitering}}, \bibinfo {author} {\bibfnamefont {J.~M.}\ \bibnamefont
  {Carpinelli}}, \bibinfo {author} {\bibfnamefont {A.~V.}\ \bibnamefont
  {Melechko}}, \bibinfo {author} {\bibfnamefont {J.}~\bibnamefont {Zhang}},
  \bibinfo {author} {\bibfnamefont {M.}~\bibnamefont {Bartkowiak}}, \ and\
  \bibinfo {author} {\bibfnamefont {E.~W.}\ \bibnamefont {Plummer}},\ }\href
  {\doibase 10.1126/science.285.5436.2107} {\bibfield  {journal} {\bibinfo
  {journal} {Science}\ }\textbf {\bibinfo {volume} {285}},\ \bibinfo {pages}
  {2107} (\bibinfo {year} {1999})},\ \Eprint
  {http://arxiv.org/abs/http://science.sciencemag.org/content/285/5436/2107.full.pdf}
  {http://science.sciencemag.org/content/285/5436/2107.full.pdf} \BibitemShut
  {NoStop}%
\bibitem [{\citenamefont {Pan}\ \emph {et~al.}(2000)\citenamefont {Pan},
  \citenamefont {Hudson}, \citenamefont {Lang}, \citenamefont {Eisaki},
  \citenamefont {Uchida},\ and\ \citenamefont {Davis}}]{Pan2000}%
  \BibitemOpen
  \bibfield  {author} {\bibinfo {author} {\bibfnamefont {S.~H.}\ \bibnamefont
  {Pan}}, \bibinfo {author} {\bibfnamefont {E.~W.}\ \bibnamefont {Hudson}},
  \bibinfo {author} {\bibfnamefont {K.~M.}\ \bibnamefont {Lang}}, \bibinfo
  {author} {\bibfnamefont {H.}~\bibnamefont {Eisaki}}, \bibinfo {author}
  {\bibfnamefont {S.}~\bibnamefont {Uchida}}, \ and\ \bibinfo {author}
  {\bibfnamefont {J.~C.}\ \bibnamefont {Davis}},\ }\href {\doibase
  10.1038/35001534M3} {\bibfield  {journal} {\bibinfo  {journal} {Nature}\
  }\textbf {\bibinfo {volume} {403}},\ \bibinfo {pages} {746} (\bibinfo {year}
  {2000})}\BibitemShut {NoStop}%
\bibitem [{\citenamefont {Isao}\ and\ \citenamefont
  {Masayasu}(1988)}]{Kojima1988}%
  \BibitemOpen
  \bibfield  {author} {\bibinfo {author} {\bibfnamefont {K.}~\bibnamefont
  {Isao}}\ and\ \bibinfo {author} {\bibfnamefont {K.}~\bibnamefont
  {Masayasu}},\ }\href {\doibase 10.1380/jsssj.9.621} {\bibfield  {journal}
  {\bibinfo  {journal} {Hyomen Kagaku}\ }\textbf {\bibinfo {volume} {9}},\
  \bibinfo {pages} {621} (\bibinfo {year} {1988})}\BibitemShut {NoStop}%
\bibitem [{\citenamefont {Heben}\ \emph {et~al.}(1988)\citenamefont {Heben},
  \citenamefont {Dovek}, \citenamefont {Lewis}, \citenamefont {Penner},\ and\
  \citenamefont {Quate}}]{Heben1988}%
  \BibitemOpen
  \bibfield  {author} {\bibinfo {author} {\bibfnamefont {M.~J.}\ \bibnamefont
  {Heben}}, \bibinfo {author} {\bibfnamefont {M.~M.}\ \bibnamefont {Dovek}},
  \bibinfo {author} {\bibfnamefont {N.~S.}\ \bibnamefont {Lewis}}, \bibinfo
  {author} {\bibfnamefont {R.~M.}\ \bibnamefont {Penner}}, \ and\ \bibinfo
  {author} {\bibfnamefont {C.~F.}\ \bibnamefont {Quate}},\ }\href {\doibase
  10.1111/j.1365-2818.1988.tb01434.x} {\bibfield  {journal} {\bibinfo
  {journal} {Journal of Microscopy}\ }\textbf {\bibinfo {volume} {152}},\
  \bibinfo {pages} {651} (\bibinfo {year} {1988})}\BibitemShut {NoStop}%
\bibitem [{\citenamefont {Valencia}\ \emph {et~al.}(2015)\citenamefont
  {Valencia}, \citenamefont {Thaker}, \citenamefont {Derouin}, \citenamefont
  {Valencia}, \citenamefont {Farber}, \citenamefont {Gebel},\ and\
  \citenamefont {Killelea}}]{Victor2015}%
  \BibitemOpen
  \bibfield  {author} {\bibinfo {author} {\bibfnamefont {V.~A.}\ \bibnamefont
  {Valencia}}, \bibinfo {author} {\bibfnamefont {A.~A.}\ \bibnamefont
  {Thaker}}, \bibinfo {author} {\bibfnamefont {J.}~\bibnamefont {Derouin}},
  \bibinfo {author} {\bibfnamefont {D.~N.}\ \bibnamefont {Valencia}}, \bibinfo
  {author} {\bibfnamefont {R.~G.}\ \bibnamefont {Farber}}, \bibinfo {author}
  {\bibfnamefont {D.~A.}\ \bibnamefont {Gebel}}, \ and\ \bibinfo {author}
  {\bibfnamefont {D.~R.}\ \bibnamefont {Killelea}},\ }\href {\doibase
  10.1116/1.4904347} {\bibfield  {journal} {\bibinfo  {journal} {Journal of
  Vacuum Science \& Technology A: Vacuum, Surfaces, and Films}\ }\textbf
  {\bibinfo {volume} {33}},\ \bibinfo {pages} {023001} (\bibinfo {year}
  {2015})},\ \Eprint {http://arxiv.org/abs/http://dx.doi.org/10.1116/1.4904347}
  {http://dx.doi.org/10.1116/1.4904347} \BibitemShut {NoStop}%
\bibitem [{\citenamefont {Ernst}\ \emph {et~al.}(2007)\citenamefont {Ernst},
  \citenamefont {Wirth}, \citenamefont {Rams}, \citenamefont {Dolocan},\ and\
  \citenamefont {Steglich}}]{Ernst2007}%
  \BibitemOpen
  \bibfield  {author} {\bibinfo {author} {\bibfnamefont {S.}~\bibnamefont
  {Ernst}}, \bibinfo {author} {\bibfnamefont {S.}~\bibnamefont {Wirth}},
  \bibinfo {author} {\bibfnamefont {M.}~\bibnamefont {Rams}}, \bibinfo {author}
  {\bibfnamefont {V.}~\bibnamefont {Dolocan}}, \ and\ \bibinfo {author}
  {\bibfnamefont {F.}~\bibnamefont {Steglich}},\ }\href {\doibase
  10.1016/j.stam.2007.05.008} {\bibfield  {journal} {\bibinfo  {journal}
  {Science and Technology of Advanced Materials}\ }\textbf {\bibinfo {volume}
  {8}},\ \bibinfo {pages} {347} (\bibinfo {year} {2007})}\BibitemShut {NoStop}%
\bibitem [{\citenamefont {Ludoph}\ and\ \citenamefont {van
  Ruitenbeek}(2000)}]{Ludoph2000}%
  \BibitemOpen
  \bibfield  {author} {\bibinfo {author} {\bibfnamefont {B.}~\bibnamefont
  {Ludoph}}\ and\ \bibinfo {author} {\bibfnamefont {J.~M.}\ \bibnamefont {van
  Ruitenbeek}},\ }\href {\doibase 10.1103/PhysRevB.61.2273} {\bibfield
  {journal} {\bibinfo  {journal} {Physical Review B}\ }\textbf {\bibinfo
  {volume} {61}},\ \bibinfo {pages} {2273} (\bibinfo {year} {2000})},\ \Eprint
  {http://arxiv.org/abs/9908139} {arXiv:9908139 [cond-mat]} \BibitemShut
  {NoStop}%
\bibitem [{\citenamefont {Wagner}\ and\ \citenamefont
  {Temirov}(2015)}]{Wagner2015}%
  \BibitemOpen
  \bibfield  {author} {\bibinfo {author} {\bibfnamefont {C.}~\bibnamefont
  {Wagner}}\ and\ \bibinfo {author} {\bibfnamefont {R.}~\bibnamefont
  {Temirov}},\ }\href {\doibase 10.1016/j.progsurf.2015.01.001} {\bibfield
  {journal} {\bibinfo  {journal} {Progress in Surface Science}\ }\textbf
  {\bibinfo {volume} {90}},\ \bibinfo {pages} {194} (\bibinfo {year}
  {2015})}\BibitemShut {NoStop}%
\bibitem [{\citenamefont {Chen}(1990)}]{Chen1990}%
  \BibitemOpen
  \bibfield  {author} {\bibinfo {author} {\bibfnamefont {C.~J.}\ \bibnamefont
  {Chen}},\ }\href {\doibase 10.1103/PhysRevB.42.8841} {\bibfield  {journal}
  {\bibinfo  {journal} {Phys. Rev. B}\ }\textbf {\bibinfo {volume} {42}},\
  \bibinfo {pages} {8841} (\bibinfo {year} {1990})}\BibitemShut {NoStop}%
\bibitem [{\citenamefont {Chiang}\ \emph {et~al.}(2014)\citenamefont {Chiang},
  \citenamefont {Xu}, \citenamefont {Han},\ and\ \citenamefont
  {Ho}}]{Chiang2014}%
  \BibitemOpen
  \bibfield  {author} {\bibinfo {author} {\bibfnamefont {C.-l.}\ \bibnamefont
  {Chiang}}, \bibinfo {author} {\bibfnamefont {C.}~\bibnamefont {Xu}}, \bibinfo
  {author} {\bibfnamefont {Z.}~\bibnamefont {Han}}, \ and\ \bibinfo {author}
  {\bibfnamefont {W.}~\bibnamefont {Ho}},\ }\href {\doibase
  10.1126/science.1253405} {\bibfield  {journal} {\bibinfo  {journal}
  {Science}\ }\textbf {\bibinfo {volume} {344}},\ \bibinfo {pages} {885}
  (\bibinfo {year} {2014})},\ \Eprint
  {http://arxiv.org/abs/http://science.sciencemag.org/content/344/6186/885.full.pdf}
  {http://science.sciencemag.org/content/344/6186/885.full.pdf} \BibitemShut
  {NoStop}%
\bibitem [{\citenamefont {Welker}\ and\ \citenamefont
  {Giessibl}(2012)}]{Joachim2012}%
  \BibitemOpen
  \bibfield  {author} {\bibinfo {author} {\bibfnamefont {J.}~\bibnamefont
  {Welker}}\ and\ \bibinfo {author} {\bibfnamefont {F.~J.}\ \bibnamefont
  {Giessibl}},\ }\href {\doibase 10.1126/science.1219850} {\bibfield  {journal}
  {\bibinfo  {journal} {Science}\ }\textbf {\bibinfo {volume} {336}},\ \bibinfo
  {pages} {444} (\bibinfo {year} {2012})},\ \Eprint
  {http://arxiv.org/abs/http://science.sciencemag.org/content/336/6080/444.full.pdf}
  {http://science.sciencemag.org/content/336/6080/444.full.pdf} \BibitemShut
  {NoStop}%
\bibitem [{\citenamefont {Welker}, \citenamefont {Weymouth},\ and\
  \citenamefont {Giessibl}(2013)}]{Joachim2013}%
  \BibitemOpen
  \bibfield  {author} {\bibinfo {author} {\bibfnamefont {J.}~\bibnamefont
  {Welker}}, \bibinfo {author} {\bibfnamefont {A.~J.}\ \bibnamefont
  {Weymouth}}, \ and\ \bibinfo {author} {\bibfnamefont {F.~J.}\ \bibnamefont
  {Giessibl}},\ }\href {\doibase 10.1021/nn403106v} {\bibfield  {journal}
  {\bibinfo  {journal} {ACS Nano}\ }\textbf {\bibinfo {volume} {7}},\ \bibinfo
  {pages} {7377} (\bibinfo {year} {2013})},\ \bibinfo {note} {pMID: 23841516},\
  \Eprint {http://arxiv.org/abs/http://dx.doi.org/10.1021/nn403106v}
  {http://dx.doi.org/10.1021/nn403106v} \BibitemShut {NoStop}%
\bibitem [{\citenamefont {Sabater}\ \emph {et~al.}(2012)\citenamefont
  {Sabater}, \citenamefont {Untiedt}, \citenamefont {Palacios},\ and\
  \citenamefont {Caturla}}]{Sabater2012}%
  \BibitemOpen
  \bibfield  {author} {\bibinfo {author} {\bibfnamefont {C.}~\bibnamefont
  {Sabater}}, \bibinfo {author} {\bibfnamefont {C.}~\bibnamefont {Untiedt}},
  \bibinfo {author} {\bibfnamefont {J.~J.}\ \bibnamefont {Palacios}}, \ and\
  \bibinfo {author} {\bibfnamefont {M.~J.}\ \bibnamefont {Caturla}},\ }\href
  {\doibase 10.1103/PhysRevLett.108.205502} {\bibfield  {journal} {\bibinfo
  {journal} {Phys. Rev. Lett.}\ }\textbf {\bibinfo {volume} {108}},\ \bibinfo
  {pages} {205502} (\bibinfo {year} {2012})}\BibitemShut {NoStop}%
\bibitem [{\citenamefont {Agra{\"\i}t}\ \emph {et~al.}(1994)\citenamefont
  {Agra{\"\i}t}, \citenamefont {Rodrigo}, \citenamefont {Rubio}, \citenamefont
  {Sirvent},\ and\ \citenamefont {Vieira}}]{Agrait1994}%
  \BibitemOpen
  \bibfield  {author} {\bibinfo {author} {\bibfnamefont {N.}~\bibnamefont
  {Agra{\"\i}t}}, \bibinfo {author} {\bibfnamefont {J.}~\bibnamefont
  {Rodrigo}}, \bibinfo {author} {\bibfnamefont {G.}~\bibnamefont {Rubio}},
  \bibinfo {author} {\bibfnamefont {C.}~\bibnamefont {Sirvent}}, \ and\
  \bibinfo {author} {\bibfnamefont {S.}~\bibnamefont {Vieira}},\ }\href
  {\doibase 10.1016/0040-6090(94)90320-4} {\bibfield  {journal} {\bibinfo
  {journal} {Thin Solid Films}\ }\textbf {\bibinfo {volume} {253}},\ \bibinfo
  {pages} {199 } (\bibinfo {year} {1994})}\BibitemShut {NoStop}%
\bibitem [{\citenamefont {Agra{\"\i}t}, \citenamefont {Yeyati},\ and\
  \citenamefont {van Ruitenbeek}(2003)}]{Agrait2003}%
  \BibitemOpen
  \bibfield  {author} {\bibinfo {author} {\bibfnamefont {N.}~\bibnamefont
  {Agra{\"\i}t}}, \bibinfo {author} {\bibfnamefont {A.~L.}\ \bibnamefont
  {Yeyati}}, \ and\ \bibinfo {author} {\bibfnamefont {J.~M.}\ \bibnamefont {van
  Ruitenbeek}},\ }\href {\doibase 10.1016/S0370-1573(02)00633-6} {\bibfield
  {journal} {\bibinfo  {journal} {Physics Reports}\ }\textbf {\bibinfo {volume}
  {377}},\ \bibinfo {pages} {81 } (\bibinfo {year} {2003})}\BibitemShut
  {NoStop}%
\bibitem [{\citenamefont {Castellanos-Gomez}\ \emph {et~al.}(2012)\citenamefont
  {Castellanos-Gomez}, \citenamefont {Rubio-Bollinger}, \citenamefont
  {Garnica}, \citenamefont {Barja}, \citenamefont {de~Parga}, \citenamefont
  {Miranda},\ and\ \citenamefont {Agra{\"\i}t}}]{Andres2012}%
  \BibitemOpen
  \bibfield  {author} {\bibinfo {author} {\bibfnamefont {A.}~\bibnamefont
  {Castellanos-Gomez}}, \bibinfo {author} {\bibfnamefont {G.}~\bibnamefont
  {Rubio-Bollinger}}, \bibinfo {author} {\bibfnamefont {M.}~\bibnamefont
  {Garnica}}, \bibinfo {author} {\bibfnamefont {S.}~\bibnamefont {Barja}},
  \bibinfo {author} {\bibfnamefont {A.~L.~V.}\ \bibnamefont {de~Parga}},
  \bibinfo {author} {\bibfnamefont {R.}~\bibnamefont {Miranda}}, \ and\
  \bibinfo {author} {\bibfnamefont {N.}~\bibnamefont {Agra{\"\i}t}},\ }\href
  {\doibase 10.1016/j.ultramic.2012.07.021} {\bibfield  {journal} {\bibinfo
  {journal} {Ultramicroscopy}\ }\textbf {\bibinfo {volume} {122}},\ \bibinfo
  {pages} {1 } (\bibinfo {year} {2012})}\BibitemShut {NoStop}%
\bibitem [{\citenamefont {Sabater}(2013)}]{Carlos2013}%
  \BibitemOpen
  \bibfield  {author} {\bibinfo {author} {\bibfnamefont {C.}~\bibnamefont
  {Sabater}},\ }\emph {\bibinfo {title} {Theoretical and experimental study of
  electronic Transport and structure in atomic-sized contacts}},\ \href@noop {}
  {Ph.D. thesis},\ \bibinfo  {school} {Universidad de Alicante} (\bibinfo
  {year} {2013})\BibitemShut {NoStop}%
\bibitem [{\citenamefont {Lang}(1986)}]{Lang1986}%
  \BibitemOpen
  \bibfield  {author} {\bibinfo {author} {\bibfnamefont {N.~D.}\ \bibnamefont
  {Lang}},\ }\href {\doibase 10.1103/PhysRevLett.56.1164} {\bibfield  {journal}
  {\bibinfo  {journal} {Phys. Rev. Lett.}\ }\textbf {\bibinfo {volume} {56}},\
  \bibinfo {pages} {1164} (\bibinfo {year} {1986})}\BibitemShut {NoStop}%
\end{thebibliography}
\end{document}